%% file: th9712.tex
\documentstyle[epsfig,12pt]{article}
\makeatletter

\newcommand{\PRL}[1]{{\it Phys.\ Rev.\ Lett.\ }{\bf #1}}
\newcommand{\PRD}[1]{{\it Phys.\ Rev.\ }{\bf D#1}}
\newcommand{\NPB}[1]{{\it Nucl.\ Phys.\ }{\bf B#1}}
\newcommand{\PLB}[1]{{\it Phys.\ Lett.\ }{\bf B#1}}
\newcommand{\ZPC}[1]{{\it Z.\ Phys.\ }{\bf C#1}}
\newcommand{\hep}[1]{[{\it hep-ph/#1}]}

\newcommand{\mrm}[1]{\mathrm{#1}}
\renewcommand{\c}{\mrm{c}}
\renewcommand{\b}{\mrm{b}}
\renewcommand{\d}{\mrm{d}}
\newcommand{\e}{\mrm{e}}
\newcommand{\g}{\mrm{g}}

\newcommand{\q}{\mrm{q}}
\newcommand{\Q}{\mrm{Q}}

\newcommand{\Z}{\mrm{Z}^0}

\newcommand{\bbbar}{\b\overline{\mrm{b}}}
\newcommand{\QQbar}{Q\overline{Q}}
\newcommand{\qqbar}{q\overline{q}}

\newcommand{\qbar}{\overline{\mrm{q}}}
\newcommand{\Qbar}{\overline{\mrm{Q}}}

\newcommand{\mc}{m_{\c}}
\newcommand{\mb}{m_{\b}}

\newcommand{\mZ}{m_{\Z}}
\newcommand{\mQ}{m_{Q}}

\newcommand{\BR}{\mrm{Br}}

\newcommand{\LQCD}{\Lambda_{\rm{QCD}}}
\newcommand{\JP}{{\mrm{J}}/\psi}
\newcommand{\OSoct}{\langle {\cal O}_8^{\Upsilon(1S)}({}^3S_1) \rangle}

\newcommand{\alphaem}{\alpha_{\mrm{em}}}

\newlength{\abstwidth}
\begin{document}

\sloppy

\renewcommand{\arraystretch}{1.5}

\pagestyle{empty}

\begin{flushright}
CERN--TH/97--12\\ 
hep-ph/9702230
\end{flushright}
 
\vspace{\fill}
 
\begin{center}
{\Large\bf 
Quarkonium production: velocity-scaling rules\\[2ex]
and long-distance matrix elements}\\[10mm]
{\Large Gerhard A. Schuler$^a$} \\[3mm]
{\it Theory Division, CERN,} \\[1mm]
{\it CH-1211 Geneva 23, Switzerland}\\[1mm]
{ E-mail: Gerhard.Schuler@cern.ch}
\end{center}
 
\vspace{\fill}
 
\begin{center}
{\bf Abstract}\\[2ex]
\begin{minipage}{\abstwidth}
The hierarchy of long-distance matrix elements (MEs) for quarkonium production 
depends on their scaling with the velocity $v$ of the heavy quark 
in the bound state. 
Ranges for the velocities in various bound states and 
uncertainties of colour-singlet MEs are estimated in a quark-potential model. 
Different possibilities for the scaling with $v$ of the MEs  
are discussed; they depend on the actual values of $v$ and the QCD scale. 
As an application, 
$\JP$ polarization in $\e^+\e^-$ annihilation is discussed. 
The first non-perturbative estimates of colour-octet MEs are presented
and compared with phenomenological determinations. 
Finally, various predictions of prompt quarkonium production at LEP 
are compared.
\end{minipage}
\end{center}

\vspace{\fill}
\noindent
\rule{60mm}{0.4mm}

\vspace{1mm} \noindent
${}^a$ Heisenberg Fellow.

\vspace{10mm}\noindent
CERN--TH/97--12 \\
January 1997

\clearpage
\pagestyle{plain}
\setcounter{page}{1} 

\section{Introduction}  
Despite the appearance of genuine non-perturbative elements, we can 
calculate the production of a heavy quarkonium bound state 
$H(J^{PC})$: 
the cross section is given as a sum of terms 
\begin{equation}
 \sigma^H = \sum_{c=1,8}\, \sum_{n}\, 
     f_c(n)\, \langle {\cal O}_c^H(n) \rangle
\ ,
\label{factorizedcross}
\end{equation}
each factorized into two parts,
$f_c(n)$ and $\langle {\cal O}_c^H(n) \rangle$, respectively, 
describing the production of a $\Q\Qbar$ pair in a 
state $|n_c\rangle$ and the transition of this state into the quarkonium $H$.
Here $c=1$ ($c=8$) denotes a colour-singlet (colour-octet) 
$\Q\Qbar$ pair. Besides spin $S$, orbital angular 
momentum $L$, and total spin $J$ of the $\Q\Qbar$ pair, one more index,
the dimension of the operator ${\cal O}$, is required in order to
specify $n$.
The short-distance part $f_c(n)$ is calculable in perturbation theory
in $\alpha_s(\mu)$ with $\mu$ of the order of the heavy-quark mass $m$. 
For a non-relativistic system such as the heavy quarkonium, the 
heavy-quark velocity $v$ (in the quarkonium rest frame) is
the natural scale, which governs the magnitude of the long-distance 
matrix elements (MEs) $\langle {\cal O}_c^H(n) \rangle$.  
For calculations accurate to a given order in $v$, only 
a finite number of terms contribute to (\ref{factorizedcross}).

The above-outlined factorization into the short-distance physics of 
the $\Q\Qbar$ pair and the long-distance physics of bound-state formation 
is a rigorous result of non-relativistic QCD (NRQCD) 
for inclusive quarkonium decays 
in the limit of large mass $m$ of the heavy quark where the 
multipole expansion is valid\cite{BBL95}.
For the case of (inclusive) quarkonium production one needs the 
additional assumption\cite{BBL95,BFY96} that all singularities, 
which neither cancel between real and virtual contributions nor can be 
absorbed into the MEs, can be factorized into the parton distribution 
functions (PDF).  A study on an extension of the factorization approach to 
exclusive decays\cite{BKS96} has recently started.

Two facts determine which MEs actually have 
to be included for an $O(v^p)$ calculation of a given process.
First the scaling of the MEs with $v$, and second the magnitude of
the respective short-distance coefficients. The latter is determined
by the powers of $\alpha_s(m)$ and, possibly, by kinematical factors.

\section{Short-distance processes}
Let us consider a few examples of short-distance reactions. 
The lowest-order short-distance processes relevant for quarkonium 
hadroproduction are
\begin{equation}
\begin{tabular}{ c c c l c }
$\q \qbar$ & $\rightarrow$ & ${\Q\Qbar}_1(X)$ 
  & $X= {}^3S_1, {}^3D_1, \ldots$
  & $\alphaem^2$
\\
  & & ${\Q\Qbar}_8(X)$ 
  & $X= {}^3S_1, {}^3D_1, \ldots$
  & $\alpha_s^2$
\\
$\g \g$ & $\rightarrow$ & ${\Q\Qbar}_1(X)$ 
  & $X= {}^1S_0, {}^3P_{0,2},  {}^1D_2, \ldots$
  & $\alpha_s^2$
\\
     &  & ${\Q\Qbar}_8(X)$ 
  & $X= {}^1S_0, {}^1P_1, {}^3P_{0,2},  {}^{2S+1}D_J, \ldots$
  & $\alpha_s^2$
\end{tabular}
\label{hadrocross}
\end{equation}
The selection rules in (\ref{hadrocross}) are easily understood. 
Annihilation of quarks and antiquarks proceeds through one-gluon
(one-photon) exchange and is, hence, restricted to $J^P=1^{-}$ states. 
Two gluons in a colour-singlet state are charge-conjugation ($C$)-even, 
while two gluons in a colour-octet state 
can be $C$-even or $C$-odd, depending on whether the $d$ or $f$ 
$SU(3)$ structure constants are involved. 
Two-gluon production of the ${}^3P_1$ state is forbidden by virtue of the
Landau--Yan theorem, independent of colour 
(provided the incident gluons can be taken on the mass shell!).
The vanishing of the squared matrix element for 
$\g\g \rightarrow {}^3S_1 + \g$ 
(again, true for on-mass-shell gluons) seems to be an accidental cancellation 
between the two QED-like diagrams and the diagram involving the three-gluon
coupling, which appears only in $1^{-}$ production. 
It would be interesting to check whether the rates for ${}^3D_1$ production 
and the $O(v^2)$ correction to ${}^3S_1$ are non-zero. 
Quarkonium production through ${}^1P_1$ and $D$-wave intermediate states 
has not yet been included in phenomenological analyses. 

In photoproduction we encounter the following reactions up to 
and including the order $\alphaem^2\alpha_s$ or $\alphaem \alpha_s^2$:
\begin{equation}
\begin{tabular}{ l l l l l }
$\gamma \g$ & $\rightarrow$
   & ${\Q\Qbar}_8(X)$ 
   & $ {}^1S_0, {}^3P_{0,2}, {}^1D_2 \ldots$
   & $e_{\Q}^2 \alphaem \alpha_s$
\\
  &
  & ${\Q\Qbar}_1(X) + \g$ 
  & $ {}^3S_1, {}^1P_{1},  {}^3D_J, \ldots$
  & $e_{\Q}^2 \alphaem \alpha_s^2$
\\
  &
  & ${\Q\Qbar}_8(X) + \g$ 
  & $ {}^{2S+1}L_J \ldots$
  & $e_{\Q}^2 \alphaem \alpha_s^2$
\\
  &
  & ${\Q\Qbar}_8(X) + \gamma$ 
  & $ {}^3S_1, {}^1P_1, {}^3D_{J} \ldots$
  & $(e_{\Q}^2 \alphaem)^2 \alpha_s$
\\
$\gamma \q$ & $\rightarrow$ 
  & ${\Q\Qbar}_8(X) + \q$ 
  & $ {}^1S_0, {}^3P_J, {}^1D_{2} \ldots$
  & $e_{\Q}^2 \alphaem \alpha_s^2$
\\
  &
  & ${\Q\Qbar}_8(X) + \q$ 
  & $ {}^3S_1, {}^3D_{1} \ldots$
  & $e_{\q}^2 \alphaem \alpha_s^2$
\end{tabular}
\label{photocross}
\end{equation}
The selection rules for the reactions involving only one gluon 
(first and fourth ones) are the same as in QED.  The second 
reaction is again restricted by the definite $C$ property of the
two-gluon colour-singlet state. Photon--quark-initiated reactions 
differ whether the photon couples to the heavy-quark line 
($\propto e_{\Q}^2$, where $e_{\Q}$ is the electric charge 
of the heavy quark in units of $e$) or the light-quark line 
($\propto e_{\q}^2$). 

In electron--positron annihilation via the exchange of an $s$-channel 
photon or $\Z$-boson, the following final states can be reached:
\begin{equation}
\begin{tabular}{ l l l l l }
 & $(g^{\Q}_V)^2$  & $(g^{\Q}_A)^2$  & $(g^{\q}_{V,A})^2$  & 
\nonumber\\
${\Q\Qbar}_8(X) + \g$ 
  & $ {}^1S_0, {}^3P_J, {}^1D_2, \ldots$
  & $ {}^3S_1, {}^1P_1, {}^3D_J, \ldots$
  & --
  & $\alpha_s\, \xi$
\nonumber\\
${\Q\Qbar}_{1,8}(X) + \g\g$ 
  & $ {}^3S_1, {}^1P_1, {}^3D_J, \ldots$
  & $ {}^1S_0, {}^3P_J, {}^1D_2, \ldots$
  & --
  & $\alpha_s^2\, \xi$
\nonumber\\
${\Q\Qbar}_{1,8}(X) + \Q\Qbar$ 
  & ${}^{2S+1}L_J$
  & ${}^{2S+1}L_J$
  & --
  & $\alpha_s^2$
\nonumber\\
${\Q\Qbar}_{8}(X) + \q\qbar$ 
  & $ {}^1S_0, {}^3P_J, {}^1D_2, \ldots$
  & $ {}^3S_1, {}^1P_1, {}^3D_J, \ldots$
  & --
  & $\alpha_s^2\, \xi$
\nonumber\\
${\Q\Qbar}_{8}(X) + \q\qbar$ 
  & --
  & --
  & ${}^3S_1$
  & $\alpha_s^2\, \ln^2\xi$
\end{tabular}
\label{electroncross}
\end{equation}
The selection rules for one-photon exchange processes are the same 
as for the analogous photoproduction processes. The same holds 
true for the vector-current part of $\Z$ exchange ($\propto 
g_V^{\Q}$). The axial-vector current contribution has opposite 
$C$-transformation property to the vector-current contribution. 
Therefore any $\Q\Qbar(X)$ state can be reached through $\Z$ exchange. 
I have also indicated the behaviour with $\xi = (2m/\sqrt{s})^2$, where
$\sqrt{s}$ denotes the $\e^+\e^-$ c.m.\ energy. 

One immediate consequence of (\ref{electroncross}) is the value of 
$A$ in the angular distribution,  
$\d\sigma /\d \cos\theta \propto 1 + A\, \cos^2\theta$, 
of (direct) $\JP$ production with respect to the beam axis 
at large $z = 2 E_{\JP}/\sqrt{s}$ and for $m_{\JP} \ll \sqrt{s} \ll \mZ$
(CLEO energies, $\sqrt{s} \sim 10\,$GeV). 
If colour-octet production is allowed (as is the case in both the 
colour-evaporation model (CEM) and in NRQCD, see below) 
then ${}^1S_0 + \g$ production dominates, 
resulting\cite{Fritzsch80,Chen96} in $A=1$. 
Explicitly\cite{Chen96}, for $r \ll 1$
\begin{equation}
  \frac{\d\sigma}{\d\cos\theta} = \frac{16 \pi^2\alphaem^2\alpha_s}{9ms^2}
  \, \left\{ \langle {\cal O}_8^{\JP}({}^1S_0)\rangle + \frac{3}{m^2}\, 
             \langle {\cal O}_8^{\JP}({}^3P_0)\rangle \right\}
\, \left(1 + \cos^2\theta\right)
\ .
\label{eq:Chen}
\end{equation}

On the other hand, if colour-singlet production dominates then 
$(\Q\Qbar)_1({}^3S_1) \g\g$ is the leading process, resulting 
in\cite{Driesen94} $A = -(1-\xi)/(1+\xi)$. This is easily understood:
the major configuration in the colour-singlet model (CSM) has the
two gluons recoiling against the $\JP$ with about equal energy. 
For $\xi \rightarrow 0$, i.e.\ neglecting the $\JP$ mass, 
conservation of parity and angular momentum then predicts $A = -1$.
Hence, the measurement of the angular 
distribution appears as a gold-plated test of the colour-octet mechanism. 

However, also in the CSM differential cross section there is an 
$A=+1$ component, namely when the energy of one gluon is at its 
lower limit (and, necessarily, the other gluon's energy at its maximum value).
Integrating the differential cross section over the region $E_{i} \leq 
m v^2$, where $E_{i}$ are the gluon energies in the $\Q\Qbar$ rest frame, 
I find in the limit $r \ll 1$ and $v^2 \ll 1$ the NRQCD result 
(\ref{eq:Chen}) with the curly bracket replaced by
\begin{equation}
  \frac{64}{81}\, \frac{\alpha_s(mv^2)}{\pi}\, v^4\, 
   \langle {\cal O}_1^{\JP}({}^3S_1)\rangle 
\ .
\end{equation}
Thus, if physics emphasizes the soft-gluon region the CSM result  
reproduces the NRQCD one with $\alpha_s(m v^2) \sim 1$. 

Alternatively, we can assume that gluons up to energies $k \sim mv$ 
are important. Integrating the squared matrix element 
over the region $E_{i} \leq m v$, I again obtain a NRQCD-type result, 
but this time with the curly bracket replaced by a term proportional 
to $\alpha_s(mv) v^2 |R(0)|^2$. Identifying $\alpha_s(mv) \sim v$
we now predict $\langle {\cal O}_8^{\JP}({}^1S_0)\rangle$ to scale 
like $v^3$ rather than like $v^4$ 
w.r.t.\ $\langle {\cal O}_1^{\JP}({}^3S_1)\rangle$. 

\section{Velocity-scaling rules}
The velocity scaling of the MEs $\langle {\cal O}_c^H(n) \rangle 
\equiv \langle {\cal O}_c^H(d,X) \rangle$
in NRQCD is determined by the number
of derivatives in the respective operators and the number of 
electric or magnetic dipole transitions between the $\Q\Qbar$ pair
produced at short distances and the $\Q\Qbar$ pair in the asymptotic 
quarkonium. The velocity scaling for the leading MEs of the most 
prominent quarkonium states are listed in Table~\ref{tab:scaling}. 
\input{uictabel1}
These leading MEs up to and including the order $v^4$ are the ones 
used in current phenomenological estimates\cite{BFY96} 
of production rates, in particular those MEs, which are enhanced 
by the short-distance coefficients. Note, however, that there 
are non-leading MEs that are less suppressed than 
some of the MEs given in Table~\ref{tab:scaling}. For example, 
the ME $\langle {\cal O}_1^\psi(d=8,{}^3S_1) \rangle$ is suppressed 
by only $v^2$ relative to the leading ME 
$\langle {\cal O}_1^\psi(d=6,{}^3S_1) \rangle$ given in 
Table~\ref{tab:scaling}; it therefore gives rise to the first 
$O(v)$ correction to $1^{--}$ production. Even though it has 
the same short-distance factor as the leading colour-singlet ME, it
may be favoured kinematically and, therefore, result in a sizeable
contribution\cite{Schuler}.

In the limit of small velocity, the NRQCD MEs for $S$-wave states and, 
hence, predictions for their production (and decays) 
reduce to those of the CSM. 
In the CSM all colour-octet MEs are zero, while the colour-singlet 
ones are related to the quarkonium wave function at the origin 
\begin{eqnarray}
  \langle {\cal O}_1^{\psi(nS)} (6,{}^3S_1) \rangle 
   & = & 3\, \frac{N_C}{2\pi}\, \left| R_{nS}(0) \right|^2
\nonumber\\
  \langle {\cal O}_1^{\chi_{QJ}(nP)} (8,{}^3P_J) \rangle 
   & = & (2J+1)\, \frac{3 N_C}{2\pi}\, \left| R'_{nP}(0) \right|^2
\nonumber\\
  \langle {\cal O}_1^{\eta_Q(nD)} (10,{}^1D_2) \rangle 
   & = & 5\, \frac{15 N_C}{8\pi}\, \left| R''_{nD}(0) \right|^2
\ .
\label{eq:CSMNRQCD}
\end{eqnarray}
The scaling of the wave function and its derivatives at $r=0$ with $v$, 
e.g.\ $|R_S(0)|^2 \propto (mv)^3$, $|R'_P(0)|^2 \propto (mv)^5$,
hence determines the scaling of the MEs in the CSM as
shown in Table~\ref{tab:scaling}. 

Opposite to the scaling rules of the CSM are those of the 
CEM. Here any $\Q\Qbar_c(X)$ state has 
probability $v^0$ to reach any quarkonium state $H$. The only 
suppression that occurs is the one in the production of the 
$\Q\Qbar_c({}^{2S+1}L_J)$ state $\propto v^{2L}$, 
see Table~\ref{tab:scaling}.

The velocity scaling rules (VSR) of NRQCD were derived\cite{VSR} on 
the basis of 
consistency requirements of the NRQCD Lagrangian. These are certainly 
valid for $\LQCD \ll m v^2 \ll m v$, i.e.\ when $\alpha_s(mv)$ is small
enough to 
renormalize the power ultraviolet-divergent contributions perturbatively.
Alternative velocity scaling rules are possible, depending on the 
sizes of $\Lambda$ and $v$. Here $\Lambda$ is ``the typical'' scale 
of QCD, i.e.\ the one that sets the scale for higher-twist corrections, 
$\Lambda$ may range from $\LQCD$ to a typical hadronic scale 
and, hence, be as large as $1\,$GeV. 
I consider two cases. First, $m v^2$ falls below 
$\Lambda$, but $\Lambda$ is still small with respect 
to $mv$, i.e.\ $\alpha_s(mv) \ll 1$. Then the ordinary multipole expansion is 
still valid, but it is $\Lambda$ that cuts off soft gluons rather than 
the binding energy\footnote{
This possibility was suggested by Martin Beneke.},
leading to a double expansion in $v \sim \langle p \rangle / m$ and 
$\lambda = k / \langle p \rangle$, where $k \sim \Lambda$ is the 
typical momentum 
of the dynamical (non-Coulombic) gluons. The scaling rules for this
alternative scenario are also displayed in Table~\ref{tab:scaling}
(labelled VSR1). These can be argued in two ways. Either by 
considering the energy shift of the $\Q\Qbar\g$ Fock state \`{a} la
BBL\cite{BBL95} or from dimensional arguments. These tell us that electric  
and magnetic dipole transitions scale as $\Gamma(E1) \propto 
g^2 k^3 r^2$ and $\Gamma(M1) \propto g^2 k^3/m^2$, respectively. 
Hence the probabilities $\Gamma/k$ scale as $\lambda^2$ for $E1$ and 
as $\lambda^2 v^2$ for $M1$ transitions owing to $r^{-1} \sim m v$.
For $k \sim m v^2$, the standard NRQCD scaling rules are reproduced. 

A second set of velocity-scaling rules 
(VSR2 in Table~\ref{tab:scaling}) is arrived at when 
$\alpha_s(mv)$ is not small enough to allow for a perturbative treatment. 
In non-relativistic Q{\bf E}D, 
the power UV-divergent terms appearing in diagrams 
containing ``soft'' photons with energy $k \sim mv$ can be cancelled 
{\em perturbatively} by corresponding terms in the bare 
coefficients\cite{Labelle96}. 
This leaves only ``ultrasoft'' photons of energy $k \sim m v^2$ for which 
the usual counting rules of the multipole expansion hold, 
$\sim v^2$ for $E1$ and $\sim v^4$ for $M1$ transitions. 
However, for the counting rules of soft photons 
only factors of $e$ and $1/m$ enter, 
so that $\psi^\dagger (g/m) \vec{p}\cdot \vec{A} \psi$ and 
$\psi^\dagger(g/m) \vec{\sigma}\cdot \vec{B} \psi$ 
contribute to the same order. Fock states that can be reached by both E1 
and M1 transitions are suppressed by $v^3$. Since 
nothing can be said about the velocity scaling of the 
non-perturbative renormalization required in NRQ{\bf C}D for 
$\alpha_s(mv)$ not small, $v^3$ remains the scaling of spin-flip transitions. 
Since the $g\vec{A}\cdot \vec{p}$ term is not renormalized 
it still scales as $v^2$.\footnote{After completion of this work I learned 
about an erratum to Ref.\cite{BBL95}, 
in which spin-flip transitions have also been assigned the 
probabilty $v^3$. I am indebted to M.\ Beneke 
for bringing this work to my attention.}$\ \ $ 
These general findings are in agreement with the explicit estimate 
of the scaling of $\langle {\cal O}_8^{\JP}({}^1S_0)\rangle$ at the end 
of section~2, where it was found that this ME scales as $v^4$ ($v^3$)
if gluons of energies $mv^2$ ($mv$) are important. 

One should bear in mind that the velocity is not a fixed quantity for 
a given quarkonium system. The velocities of $1^{--}$ and $J^{++}$ states 
are, in general, different and such is the case for different radial
excitations. Universal velocity-scaling rules are only meaningful as long as
velocity differences are small compared to the actual velocities. 
It may even happen that the relation between $\Lambda$ and $mv$ or
$mv^2$ differs from one state to the other. 

In order to study these problems I calculate the velocity $v$, 
the momentum $p$ and the kinetic energy $T$ of the heavy quark 
for the lowest-lying $S$-, $P$-, and $D$-wave quarkonium states,  
for a class of potentials of the form 
$V(r) = \lambda\, r^\nu + V_0$. A power-like potential successfully 
describes quarkonium spectroscopy 
with\cite{Martin80} $\nu \sim 0.1$. 
Uncertainties of the quark-potential model description can be investigated 
by studying the power-dependence within a reasonable range, say 
$\Delta \nu = \pm 0.2$. For every value of $\nu$ I use the 
$1S$ leptonic width and the $1S$ and $2S$ masses to fix the three 
remaining parameters $m$, $\lambda$, and $V_0$. Note that I 
normalize the leptonic width by the quarkonium mass rather than the 
heavy-quark mass and do not include the known $O(\alpha_s)$ correction.

\begin{figure}[p]
\begin{center}
\begin{tabular}{cc}
   \epsfig{file=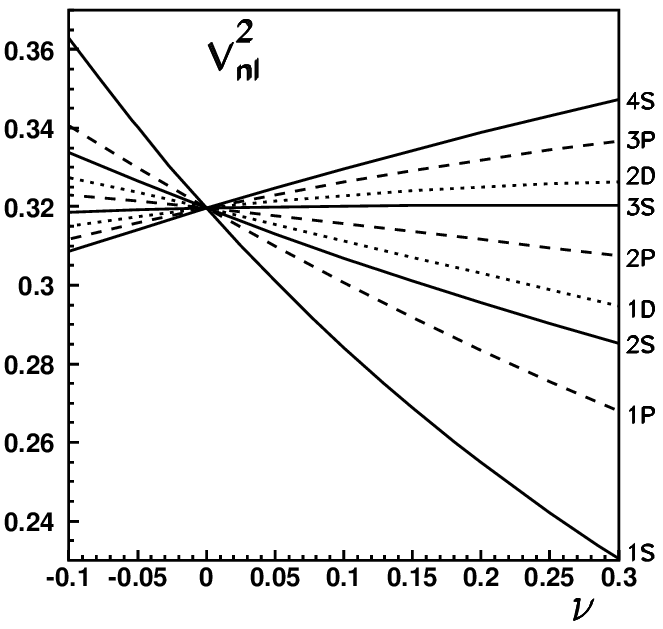,width=0.45\textwidth} &
   \epsfig{file=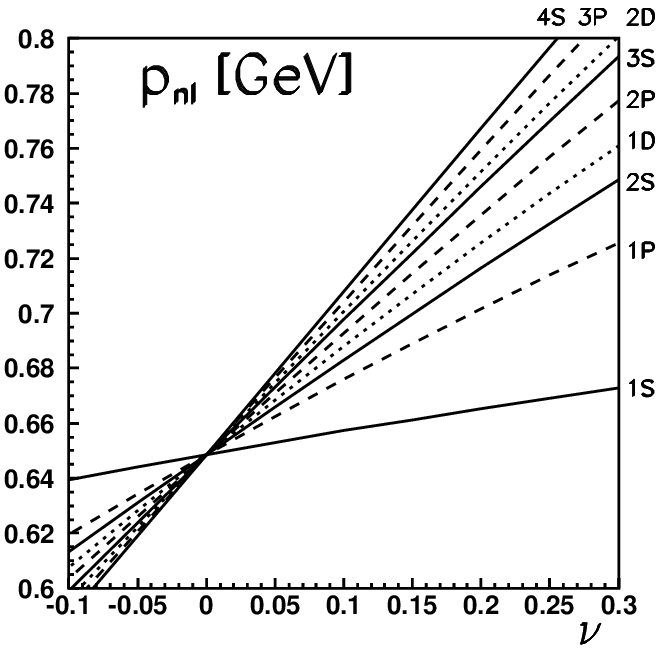,width=0.45\textwidth}
 \\
   \epsfig{file=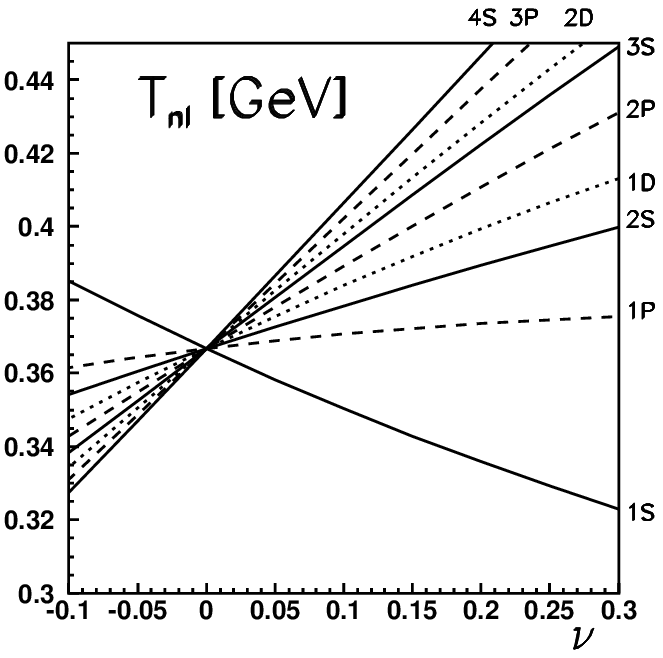,width=0.45\textwidth} &
   \epsfig{file=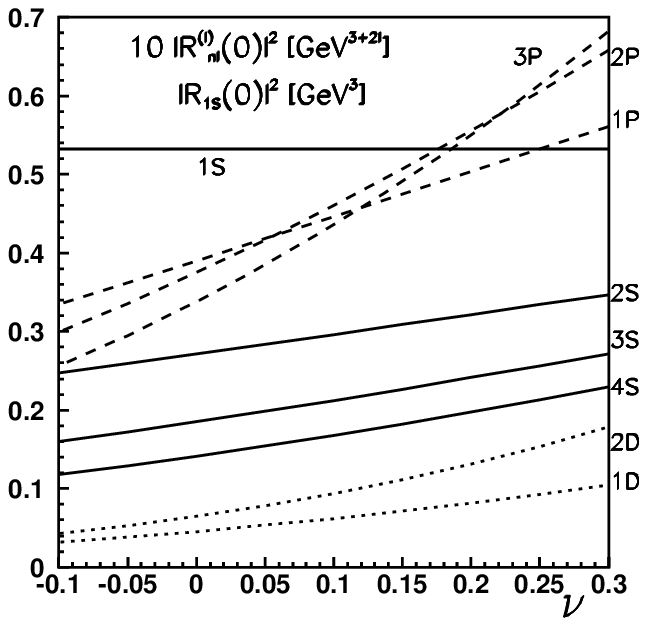,width=0.45\textwidth}
  \end{tabular}
\end{center}
\caption[]{Average squared velocity, momentum, and kinetic energy 
of the charm quark as functions of $\nu$ for charmonium states: 
$S$-waves (solid lines), $P$-waves (dashed lines), $D$-waves (dotted lines). 
Also shown is $|R_{nl}^{(l)}(0)|^2 = \d^l R_{nl}(r)/\d r^l|_{r=0}$. 
\label{fig:charm}}
\setlength{\unitlength}{1pt}
\end{figure}

\begin{figure}[p]
\begin{center}
\begin{tabular}{cc}
   \epsfig{file=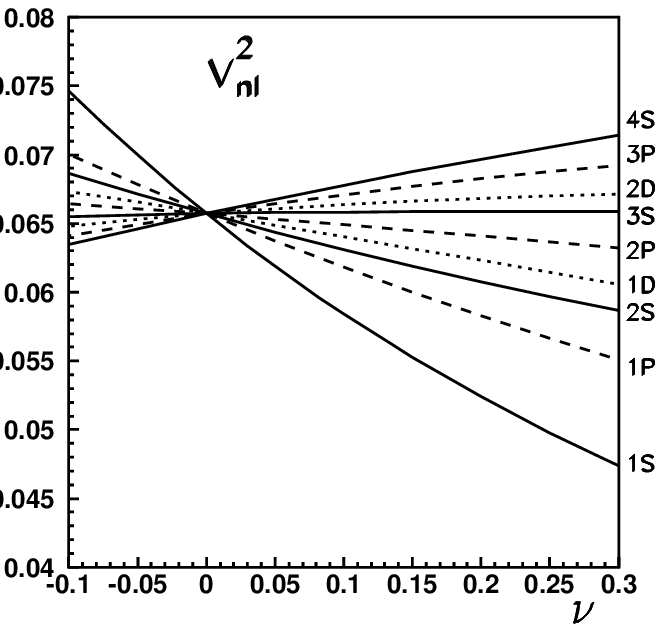,width=0.45\textwidth} &
   \epsfig{file=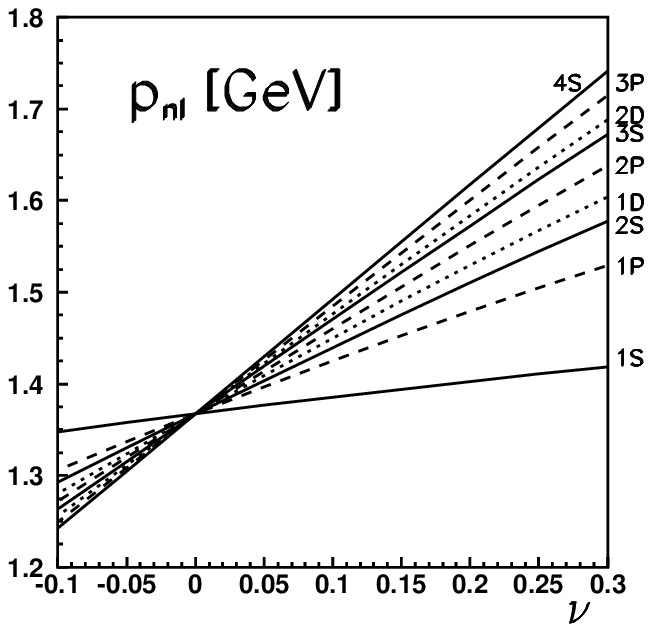,width=0.45\textwidth}
 \\
   \epsfig{file=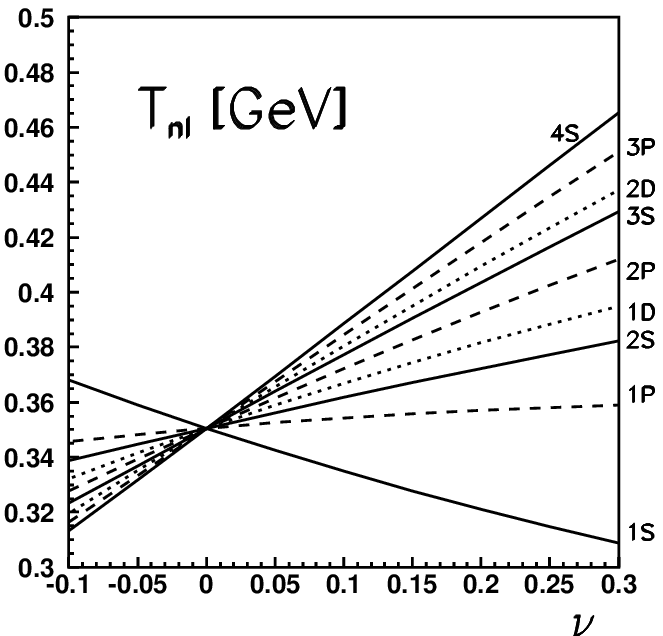,width=0.45\textwidth} &
   \epsfig{file=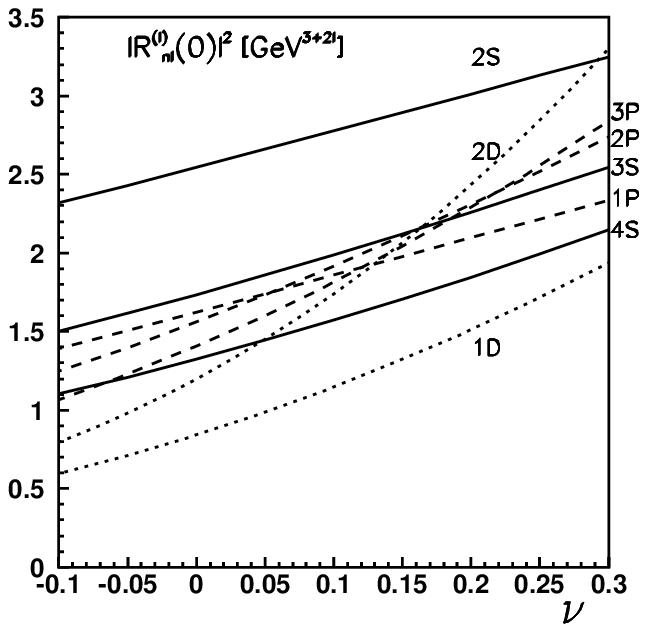,width=0.45\textwidth}
  \end{tabular}
\end{center}
\caption[]{Same as Fig.~\ref{fig:charm} but for bottomonia.
\label{fig:bottom}}
\setlength{\unitlength}{1pt}
\end{figure}

The results given in Figs.~\ref{fig:charm} and~\ref{fig:bottom} show that 
the velocities are rather uncertain: varying $\nu$ between $-0.1$ and $0.3$ 
changes, e.g.\ $v_{\JP}^2$ from $0.36$ to $0.23$. An additional uncertainty 
is caused by the $O(\alpha_s)$ correction $K = 1 - (16/3) \alpha_s/\pi$
to the leptonic width, which enters as $K^{2/3} \sim 0.63$ for charm
and $\sim 0.76$ for bottom. It can also be seen that the difference in 
the $1S$ and $2S$ squared velocities is non-negligible, it can be as large 
as $20$\%. 
Moreover, for charmonia we observe $v^2 \sim v/2 \sim \alpha_s(m)$. 
The kinetic energy $T$ is about $350\,$MeV for both the charm and the bottom 
systems. For bottomonia, however, $v^2$ is small enough for the 
NRQCD relation $v \sim \alpha_s(mv)$ to be fulfilled. 
Hence, bottomonia seem to be heavy enough for the NRQCD velocity-scaling 
rules to hold, while the charm-quark mass is possibly too light for the 
notion of a single velocity to make sense and a universal scaling 
to hold for all charmonium states. 

\section{NRQCD matrix elements}
The long-distance MEs can either be calculated using non-perturbative 
methods or extracted phenomenologically from data. 
In order to estimate the total $1^{--}$ production rates we need also 
total production MEs, which include the feed down from higher states. 
For $\Upsilon(1S)$ the total MEs are defined as follows:
\begin{eqnarray}
\left. \OSoct\right|_{\mrm tot} & = & 
\sum_{n=1}^3\,  \left\{ 
  \langle {\cal O}_8^{\Upsilon(nS)}({}^3S_1)\rangle \, 
\BR[ \Upsilon(nS) \rightarrow \Upsilon(1S)\, X]
\right. 
\nonumber\\
& &~ \left. 
 +  \sum_{J=0}^2\,
\langle {\cal O}_8^{\chi_{\b J}(nP)}({}^3S_1) \rangle \,
\BR[ \chi_{\b J}(nP)
   \rightarrow \Upsilon(1S)\, X] \right\}
\nonumber\\
 \left. \langle 
{\cal O}_1^{\Upsilon(1S)}({}^3S_1) 
  \rangle \right|_{\mrm tot} & = & 
\sum_{n=1}^3\, \langle {\cal O}_1^{\Upsilon(nS)}({}^3S_1)\rangle \, 
\BR[ \Upsilon(nS) \rightarrow \Upsilon(1S)\, X]
\ .
\label{replace}
\end{eqnarray}
The inclusive branching ratios needed for (\ref{replace}) 
and the other $\Upsilon(nS)$ states are compiled in Table~\ref{tab:BRs}.
\begin{table}[htbp]
\centerline{
\begin{tabular}{ r  r r r r r }
\hline
 & \multicolumn{2}{ c }{}
 & $\Upsilon(3S)$
 & $\Upsilon(2S)$
 & $\Upsilon(1S)$
\\ 
$\chi_{\b 2}(3P)$ 
 & \multicolumn{2}{ c }{}
 & $16.2$
 & $5.4$
 & $5.4$
\\ 
$\chi_{\b 1}(3P)$ 
 & \multicolumn{2}{ c }{} 
 & $21~~$
 & $7~~$
 & $7~~$
\\ 
$\chi_{\b 0}(3P)$ 
 & \multicolumn{2}{ c }{}
 & $~4.6$
 & $1.5$
 & $1.5$
\\ \hline
 & $\chi_{\b 2}(2P)$ & $\chi_{\b 1}(2P)$ & $\chi_{\b 0}(2P)$
 & $\Upsilon(2S)$
 & $\Upsilon(1S)$
\\ 
$\Upsilon(3S)$
 & $11.4 \pm 0.8$ & $11.3 \pm 0.6$ & $~5.4 \pm 0.6$ 
 & $10.6 \pm 0.8$
 & $11.7 \pm 0.5$
\\ 
$\chi_{\b 2}(2P)$ 
 & & & 
 & $16.2 \pm 2.4$
 & $12.~ \pm 1.3$
\\ 
$\chi_{\b 1}(2P)$ 
 & & & 
 & $21.~ \pm 4.~$
 & $15.~ \pm 1.9$
\\ 
$\chi_{\b 0}(2P)$ 
 & & & 
 & $~4.6 \pm 2.1$
 & $~2.3 \pm 0.9$
\\ \hline
 & $\chi_{\b 2}(1P)$ & $\chi_{\b 1}(1P)$ & $\chi_{\b 0}(1P)$ 
 & $\Upsilon(1S)$  
 & 
\\ 
$\Upsilon(2S)$
 & $~6.6 \pm 0.9$ & $~6.7 \pm 0.9$ & $~4.3 \pm 1.0$ 
 & $31.1 \pm 1.6$
 &
\\ 
$\chi_{\b 2}(1P)$ 
 & & & 
 & $22.~ \pm 4.~$
 & 
\\ 
$\chi_{\b 1}(1P)$ 
 & & & 
 & $35.~ \pm 8.~$
 & 
\\ 
$\chi_{\b 0}(1P)$ 
 & & & 
 & $ < 6.~ \quad$
 &
\\ \hline 
\end{tabular}}
\caption[]{Branching ratios (in per cent) in the $\bbbar$ system; 
sequential decays have been included for the $\Upsilon(3S)$, 
$\Upsilon(2S)$, and $\chi_{\b J}(2P)$ branching ratios; 
based on the PDG data\cite{PDG} and, for the $\chi_{\b J}(3P)$, 
on Ref.\cite{Grant93}.
\label{tab:BRs}}
\end{table}%

\subsection{Non-perturbative methods}
Since production MEs cannot be determined in lattice gauge theory
one has to resort to less rigorous non-perturbative models. 
The leading-order (in $v$) colour-singlet 
MEs can be estimated in the quark-potential model since they are simply 
related to the wave functions at the origin, 
see (\ref{eq:CSMNRQCD}). 
Tables~\ref{charmsingletME} and~\ref{bottomsingletME} list the values 
obtained\cite{Eichten96} from a QCD-motivated potential\cite{BT81}.
\begin{table}[htbp]
\centerline{
\begin{tabular}{ l l l l }
\hline
$\langle {\cal O}_1^{J/\psi}({}^3S_1) \rangle$
 & $1.16\,$GeV$^3$ 
& $\left. \langle {\cal O}_1^{J/\psi}({}^3S_1) \rangle \right|_{\mrm tot}$
 & $1.59\,$GeV$^3$ 
\\
$\langle {\cal O}_1^{\chi_{\c 1}}({}^3P_1) \rangle$
 & $0.32\,$GeV$^5$ 
& $\langle {\cal O}_1^{\psi(2S)}({}^3S_1) \rangle$
 & $0.76\,$GeV$^3$ 
\\ \hline
\end{tabular}}
\caption[]{Colour-singlet long-distance matrix elements for charmonium 
production
\label{charmsingletME}
}
\end{table}%
\begin{table}[htbp]
\centerline{
\begin{tabular}{ l l l l }
\hline
$\langle {\cal O}_1^{\Upsilon(1S)}({}^3S_1) \rangle$
 & $9.28\,$GeV$^3$ 
& 
$\left. \langle {\cal O}_1^{\Upsilon(1S)}({}^3S_1) \rangle \right|_{\mrm tot}$
 & $11.1\,$GeV$^3$ 
\\
$\langle {\cal O}_1^{\chi_{\b 1}(1P)}({}^3P_1) \rangle$
 & $6.09\,$GeV$^5$ 
 & &
\\
$\langle {\cal O}_1^{\Upsilon(2S)}({}^3S_1) \rangle$
 & $4.63\,$GeV$^3$ 
&
$\left. \langle {\cal O}_1^{\Upsilon(2S)}({}^3S_1) \rangle \right|_{\mrm tot}$
 & $5.01\,$GeV$^3$ 
\\
$\langle {\cal O}_1^{\chi_{\b 1}(2P)}({}^3P_1) \rangle$
 & $7.10\,$GeV$^5$ 
 & & 
\\
$\langle {\cal O}_1^{\Upsilon(3S)}({}^3S_1) \rangle$
 & $3.54\,$GeV$^3$ 
& 
$\left. \langle {\cal O}_1^{\Upsilon(3S)}({}^3S_1) \rangle \right|_{\mrm tot}$
 & $3.54\,$GeV$^3$ 
\\
$\langle {\cal O}_1^{\chi_{\b 1}(3P)}({}^3P_1) \rangle$
 & $7.71\,$GeV$^5$ 
& & 
\\ \hline
\end{tabular}}
\caption[]{Colour-singlet long-distance matrix elements for 
bottomonium production
\label{bottomsingletME}
}
\end{table}%
Uncertainties in the wave functions arise from several sources: 
relativistic corrections $\propto v^2$, the form of the potential, and 
badly determined input parameters. Among the latter the $O(\alpha_s)$ 
correction to the leptonic width gives by far the biggest uncertainty: 
the factor $1 - (16/3) \alpha_s/\pi$ can change the square of charmonium 
wave functions by as much as a factor of $2$! 

In order to assess the size of the other
effects I calculate the wave functions for a power-like potential 
$V(r) \propto \lambda\, r^\nu$ and study them 
as functions of $\nu$. In this way we modify not only the form 
of the potential but also change the velocity $v$ and, hence, 
the size of relativistic corrections. 
The $1S$ leptonic width and the $1S$ and $2S$ masses 
are used as input. The results for 
$|R_{nl}^{(l)}(0)|^2 = \d^l R_{nl}(r)/\d r^l|_{r=0}$ 
are shown in Figs.~\ref{fig:charm} and~\ref{fig:bottom}. 
In the case of charmonia, the $nS$ to $1S$ ratio is rather stable but 
the ratio $nP$ to $1S$ is quite uncertain, by a factor of almost $2$. 
This uncertainty is not much smaller for the $b$ system, in particular 
for the higher radial excitations. The $D$-wave wave functions are 
mostly uncertain. This underlines that the uncertainties
of colour-singlet MEs must not be forgotten when estimating the ranges of 
theoretical predictions for quarkonium production. 

\begin{table}[htbp]
\centerline{
\begin{tabular}{ c c l l c l l }
\hline
 & \multicolumn{3}{c}{$n=1$}  & \multicolumn{3}{c}{$n=2$} 
\\ 
 \multicolumn{2}{r}{Condensate\cite{Hoodbhoy}}
 & \multicolumn{1}{c}{CEM}
 & \multicolumn{1}{c}{Tevatron\cite{CL96}}
 & \multicolumn{1}{c}{Condensate\cite{Hoodbhoy}}
 & \multicolumn{1}{c}{CEM}
 & \multicolumn{1}{c}{Tevatron\cite{CL96}}
\\
$\langle {\cal O}_8^{\psi(nS)}({}^3S_1) \rangle$
 &
 & $14.$
 & $~6.6$ 
 &
 & $3.4$
 & $4.6$
\\
$\left. \langle {\cal O}_8^{\psi(nS)}({}^3S_1) \rangle \right|_{\mrm tot}$
 &
 & $24.$
 & $14.$
 &
 & $3.4$
 & $4.6$
\\
$\langle {\cal O}_8^{\psi(nS)}({}^1S_0) \rangle$
 & $~~5.6$
 & $~2.6$
 & $33.$
 & $0.5$
 & $0.62$
 & $8.8$
\\
$\langle {\cal O}_8^{\psi(nS)}({}^3P_0) \rangle$
 & $-0.2$
 & $~1.2$
 & $24.$
 & 
 & $0.29$
 & $6.5$
\\
$\langle {\cal O}_8^{\chi_{\c 1}}({}^3S_1) \rangle$
 & 
 & $14.$
 & $9.8$
 & & &
\\ \hline
\end{tabular}}
\caption[]{Colour-octet long-distance matrix elements 
$\langle {\cal O}_8^H({}^{2S+1}L_J) \rangle$ 
for charmonium production in units of $10^{-3}\,$GeV$^{3+2L}$. 
\label{charmoctetME}
}
\end{table}%
Colour-octet production MEs have so far been estimated only 
phenomenologically. Here I present the first non-perturbative 
estimates obtained in two different models. 
One calculation of the colour-octet ME, 
in collaboration with P.\ Hoodbhoy\cite{Hoodbhoy}, 
is based on the method of Leutwyler--Voloshin\cite{Leutwyler}. 
Here colour-octet MEs are given in terms of the gluon condensate 
$\langle \alpha_s\, G^2 \rangle$. We obtain
\begin{eqnarray}
  \langle {\cal O}_8^{\JP}({}^1S_0) \rangle & = & 
  5.6 \times 10^{-3}\, {\mrm{GeV}}^3 \,\, \left[
  \frac{ \langle \alpha_s\, G^2 \rangle }{0.1 \,{\mrm{GeV}}^4} \, 
  \frac{1.5\,{\mrm{GeV}} }{m} \,
  \frac{0.26}{\alpha_s} \right]
\nonumber\\
  \langle {\cal O}_8^{\JP}({}^3P_0) \rangle & = & - \frac{12}{625}\, 
  m^2 \,   \langle {\cal O}_8^{\JP}({}^1S_0) \rangle 
\ .
\end{eqnarray}
Note that $\langle {\cal O}_8^{\JP}({}^3P_0)\rangle$ is small and, 
as a consequence of $\langle B^2 \rangle = - \langle E^2 \rangle$, negative.

\begin{table}[tbp]
\centerline{
\begin{tabular}{ c l l l l l }
\hline
 & \multicolumn{1}{c}{Condensate\cite{Hoodbhoy}}
 & \multicolumn{1}{c}{CEM$0$}
 & \multicolumn{1}{c}{CEM$1$}
 & \multicolumn{1}{c}{CEM$100$}
 & \multicolumn{1}{c}{Tevatron\cite{CL96}}
\\
$\langle {\cal O}_8^{\Upsilon(1S))}({}^3S_1) \rangle$
 &
 & $144.$
 & $~94.$ 
 & $~~2.8$
 & $5.9$
\\
$\left. \langle {\cal O}_8^{\Upsilon(1S)}({}^3S_1) \rangle \right|_{\mrm tot}$
 &
 & $228.$
 & $228.$
 & $228.$
 & $480.$
\\
$\langle {\cal O}_8^{\Upsilon(1S)}({}^1S_0) \rangle$
 & $~~2.3$ 
 & $41.$
 & $27.$
 & $~0.80$
 & $20.$
\\
$\langle {\cal O}_8^{\Upsilon(1S)}({}^3P_0) \rangle$
 & $-1.1$
 & $55.$
 & $36.$
 & $~1.1$
 & $94.$
\\
$\langle {\cal O}_8^{\chi_{\b 1}(1P)}({}^3S_1) \rangle$
 & 
 & $~24.$
 & $~79.$
 & $189.$ 
 & $420.$ 
\\ \hline
$\langle {\cal O}_8^{\Upsilon(2S))}({}^3S_1) \rangle$
 &
 & $~55.$
 & $~36.$ 
 & $~~1.0$
 & $4.1$
\\
$\left. \langle {\cal O}_8^{\Upsilon(2S)}({}^3S_1) \rangle \right|_{\mrm tot}$
 &
 & $119.$
 & $119.$
 & $119.$
 & $220.$
\\
$\langle {\cal O}_8^{\Upsilon(2S)}({}^1S_0) \rangle$
 & 
 & $15.$
 & $10.$
 & $~0.29$
 & $23.$
\\
$\langle {\cal O}_8^{\Upsilon(2S)}({}^3P_0) \rangle$
 & 
 & $21.$
 & $14.$
 & $~0.40$
 & $110.$
\\
$\langle {\cal O}_8^{\chi_{\b 1}(2P)}({}^3S_1) \rangle$
 & 
 & $129.$
 & $159.$
 & $219.$ 
 & $320.$ 
\\ \hline
$\langle {\cal O}_8^{\Upsilon(3S))}({}^3S_1) \rangle$
 &
 & $~39.$
 & $~26.$ 
 & $~~0.75$
 & $4.1$
\\
$\left. \langle {\cal O}_8^{\Upsilon(3S)}({}^3S_1) \rangle \right|_{\mrm tot}$
 &
 & $~39.$
 & $~39.$
 & $~39.$
 & $160.$
\\
$\langle {\cal O}_8^{\Upsilon(3S)}({}^1S_0) \rangle$
 & 
 & $11.$
 & $~7.3$
 & $~0.21$
 & $23.$
\\
$\langle {\cal O}_8^{\Upsilon(3S)}({}^3P_0) \rangle$
 & 
 & $15.$
 & $~9.9$
 & $~0.28$
 & $110.$
\\
$\langle {\cal O}_8^{\chi_{\b 1}(3P)}({}^3S_1) \rangle$
 & 
 & $~~0.$
 & $~25.$
 & $~75.$ 
 & $320.$ 
\\ \hline
$\langle {\cal O}_8^{\sum \Upsilon}({}^3S_1) \rangle$
 &
 & $387.$
 & $387.$
 & $387.$
 & $850.$
\\ \hline
\end{tabular}}
\caption[]{Colour-octet long-distance matrix elements 
$\langle {\cal O}_8^H({}^{2S+1}L_J) \rangle$ 
for bottomonium production in units of $10^{-3}\,$GeV$^{3+2L}$. 
\label{bottomoctetME}
}
\end{table}%
Long-distance MEs can also be estimated by comparing NRQCD calculations 
with CEM ones. Leaving the details to a further publication 
I simply quote the results
\begin{eqnarray}
 \left.  \langle {\cal O}_8^{H}({}^3S_1) \rangle\right|_{\mrm{dir}}
    & = & 
     \left. F[H]\right|_{\mrm{dir}}\, \frac{4}{\pi^2}\,
         \left( 2\, \mQ\, \bar{\Lambda} 
            + \bar{\Lambda}^2\right)^{3/2}
\nonumber\\
     \langle {\cal O}_8^{H}({}^1S_0) \rangle 
   & = & \frac{1}{3 + 9\, v^2}\, 
   \left. \langle {\cal O}_8^{H}({}^3S_1) \rangle\right|_{\mrm{dir}} 
\nonumber\\
         \langle {\cal O}_8^{H}({}^3P_0) \rangle 
   & = & m^2\, v^2\, 
    \langle {\cal O}_8^{H}({}^1S0) \rangle
\ .
\end{eqnarray}
Here $\bar{\Lambda} = m(H_{\Q}) - \mQ$ is the difference between
the mass of the heavy quark and that of the lightest meson containing it. 
The long-distance factors $F[H]$ can be extracted 
from the ratios of the production rates for the various $\Upsilon(nS)$ 
($\psi(2S)$, $\JP$) states known from fixed-target experiments 
together with spin symmetry and an ansatz for direct $\Upsilon(nS)$ 
production, which relates these to the respective leptonic widths. 
The results are given in Tables~\ref{charmoctetME}
and~\ref{bottomoctetME}. In the case of bottomonia, I present my  
estimates as a function of the size of the cross section of the as yet 
unobserved $3P$ $\chi_{\b J}$ states. The models CEM$0$, CEM$1$, and
CEM$100$ are defined by $R_{3P}  = 0$, $1$, and $100$, respectively, where
\begin{equation}
  R_{3P} = \frac{\sigma[\chi_{\b 1}(3P)]}
                {\sigma_{\mrm{dir}}[\Upsilon(3S)]}
\ . 
\label{eq:R3P}
\end{equation} 
We consider CEM$1$ as the most sensible choice. 
  
\subsection{Phenomenological determination of MEs}
Tables~\ref{charmoctetME} and~\ref{bottomoctetME} quote 
also the values of colour-octet MEs determined in a 
study of quarkonium production at the Tevatron\cite{CL96}. 
The usual assumption 
$m^2 \langle {\cal O}_8^H({}^1S_0)\rangle 
= 3 \langle {\cal O}_8^H({}^3P_0)\rangle$ 
is made and, for bottomonium, I have identified 
the $3L$ MEs with the corresponding $2L$ ones ($L=S,P$).

The values of phenomenological determinations of (in particular
colour-octet) MEs should be considered with great care. Numbers 
found in different processes may deviate by large factors 
and, yet, be consistent if proper error estimates were indicated.
Uncertainties arise from several sources. 
First, a substantial uncertainty is caused by the 
large value of $\alpha_s(\mu)$, which enters the various processes 
with different powers. Realistic error estimates should take into account
the variations of $\LQCD$ and the scale $\mu \sim \sqrt{m^2 + p_T^2}$. 

A second source of uncertainty is introduced by the 
parton distribution functions (PDFs). Quark and gluon distribution functions 
enter in different combinations, at different scales, and in different
$x$ regions. Hence, comparisons of $B$ decays into charmonium 
(independent of PDFs), photoproduction of $J/\psi$ at low energies 
(dominated by single parton-distributions at large $x$), 
and $J/\psi$ production at the Tevatron 
(governed by the product of two gluon-distributions at small $x$) 
are highly non-trivial. 

Third, essentially all analyses are leading order in both $v$ 
and $\alpha_s(m)$ but, the size of higher-order corrections in $v$ and/or  
$\alpha_s$ may well be process-dependent. Moreover, it should be stressed 
that existing calculations are separately of leading order in $v$ for 
the colour-singlet and colour-octet contributions, i.e.\ 
do not consistently include all terms of a given order in $v$.

Last but not least, 
there exist higher-twist corrections, which enter at different levels.
Examples are diffractive contributions to the $z \rightarrow 1$ behaviour
of $J/\psi$ photoproduction $\propto \Lambda^2/(1-z)\mc^2$, 
$\Lambda/(1-x_F)\mc^2$ effects at large-$x_F$ 
charmonium production, and effects of intrinsic $k_T$ at low-$p_T$ 
quarkonium production at the Tevatron. 

\section{Discussion}
\noindent
We start by comparing the values of the charmonium colour-octet MEs 
extracted from the Tevatron with the non-perturbative results. The MEs 
$\langle {\cal O}_8^H({}^3S_1) \rangle$ 
come out quite similar for $H=\JP$, $\psi(2S)$, and $\chi_{\c J}$. 
This is not surprising since ratios of production cross sections for 
these states $H$ are rather universal\cite{Our} and, hence, 
the MEs are well constrained. 
On the other hand, where there are less data, differences show up: 
the Tevatron values for 
$\langle {\cal O}_8^H({}^1S_0,{}^3P_0) \rangle$ 
are considerably larger than the non-perturbative ones. 

Turning now to bottomonia, we find similar values for
$\langle {\cal O}_8^\Upsilon({}^1S_0) \rangle$ 
if $R_{3P}$ (see (\ref{eq:R3P})) 
is close to our preferred value $\sim 1$. This also holds for the 
actual combination extracted from the Tevatron, 
$\langle {\cal O}_8^\Upsilon({}^1S_0) \rangle/3 + 
\langle {\cal O}_8^\Upsilon({}^3P_0) \rangle/\mb^2$, 
which was found\cite{CL96} to be $7.9$, to be compared with 
$16$, $10$, $0.31$ ($\times 10^{-3}\,$GeV$^3$)
for $R_{3P}=0$, $1$, $100$, respectively. 

For the choice $R_{3P}=1$, however, the Tevatron value for 
$\langle {\cal O}_8^\chi({}^3S_1) \rangle$ 
is much larger than the CEM1 one. On the other hand, the situation 
is reversed for 
$\langle {\cal O}_8^\Upsilon({}^3S_1) \rangle|_{\mrm{dir}}$ 
so that, in fact, the values for 
$\langle {\cal O}_8^\Upsilon({}^3S_1) \rangle|_{\mrm{tot}}$ 
are rather similar. Again not a surprise, given that so far only 
total (direct plus indirect) $\Upsilon(nS)$ cross sections have been
measured. Since the Tevatron values correspond to very large 
production cross sections for the as yet unobserved $\chi_{\b J}(3P)$ 
states, it is likely that the ${}^3S_1$ octet MEs for $\chi_{\b J}(nP)$ 
are overestimated while the ones for $\Upsilon(nS)$ are underestimated.

\begin{table}[htbp]
\centerline{
\begin{tabular}{ r | l l | l l | l l | l  }
\hline
 \multicolumn{1}{ c }{}
 & \multicolumn{2}{ c }{$\Z \rightarrow \left(\QQbar\right)_8 \g$}
 & \multicolumn{2}{ c }{$\Z \rightarrow \left(\QQbar\right)_8 \qqbar$}
 & \multicolumn{2}{ c }{$\Z \rightarrow \Q \left(\QQbar\right)_1 \Qbar$}
 & \multicolumn{1}{ c }{Data}
\\ 
 & CEM & NRQCD & CEM & NRQCD & CEM & NRQCD & 
\\ 
$\left. J/\psi\right|_{\mrm{dir}}$       
 & $0.014$  & $0.023$ 
 & $13.6$    & $~8.4$ 
 & $0.14$    & $5.4$
 &
\\ 
$\left. J/\psi\right|_{\mrm{tot}}$       
 & $0.023$     & $0.028$  
 & $22.7$       & $18$ 
 & $0.23$      & $7.8$ 
 & $19\pm 10$ \cite{LEPpsi}
\\ 
$\psi(2S)$
 & $0.0033$ & $0.006$  
 & $~3.3$   & $~5.9$ 
 & $0.033$   & $3.6$ 
 &
\\ 
$\chi_{\c 0}$
 & $0.0046$ & $0.002$  
 & $~4.5$   & $~4.2$ 
 & $0.046$   & $0.99$ 
 &
\\ 
$\chi_{\c 1}$
 & $0.014$ & $0.006$  
 & $13.6$   & $12.5$ 
 & $0.14$   & $1.13$ 
 &
\\ 
$\chi_{\c 2}$
 & $0.023$ & $0.010$  
 & $22.7$   & $20.9$ 
 & $0.23$   & $0.31$ 
 &
\\ 
$\left. \Upsilon(1S)\right|_{\mrm{dir}}$ 
 & $0.030$  & $0.0076$
 & $~0.60$ & $~0.034$
 & $0.014$  & $0.76$
 &
\\ 
 & $0.020$  & 
 & $~0.39$  &
 & $0.0090$ &
 &
\\ 
 & $0.00059$  & 
 & $~0.012$ &
 & $0.00027$  &
 &
\\ 
$\sum \Upsilon$
 & $0.081$   & $0.14$ 
 & $~1.6$  & $~4.8$
 & $0.037$   & $1.7$  
 & $10 \pm 5$ \cite{LEPups} 
\\ \hline 
\end{tabular}}
\caption[]{Branching ratios (in units of $10^{-5}$) of $\Z$ decays
\label{Zdecaytable}
}
\end{table}%
Finally, I consider prompt quarkonium production from $\Z$ decays at LEP 
and calculate branching ratios in both NRQCD and the CEM. 
I improve previous calculations in the 
CEM\cite{Branco,Kane,Fritzsch80,Eboli96} and the 
CSM\cite{Kuhn81,Clavelli82,Braaten93,Driesen94}%
/NRQCD\cite{Cho95} 
by carefully including in the $\JP$ and $\sum\Upsilon$ 
cross sections the feed down from higher states. Moreover, branching ratios 
for various other states are also presented. 
Also the CEM prediction for the $(\Q\Qbar)_1+\Q\Qbar$ final state is 
calculated here for the first time. The results (Table~\ref{Zdecaytable}) 
are obtained with the help of the MEs given in  
Tables~\ref{charmsingletME}--\ref{bottomoctetME}. For direct 
$\Upsilon(1S)$ production, estimates corresponding to $R_{3P}=0$,
$1$, and $100$ are given. 

It can be seen that NRQCD and CEM predictions 
are rather similar for the colour-octet processes, but differ 
substantially for the colour-singlet one. Both predictions are compatible 
with the data. Concerning the NRQCD predictions: one can see that the ratio 
of colour-singlet to colour-octet production decreases 
when going from $J/\psi$ to the sum of the $\Upsilon$ states. 
This is in contrast with the expectation, since the octet contribution 
scales as $v^4 \ln^2 \xi$ ($\sqrt{\xi}=2m_{\Q}/\sqrt{s}$) and both factors 
are larger for charmonium. This supports the observation made above that the
colour-octet MEs of the $P$-wave bottomonium states as 
obtained\cite{CL96} from the Tevatron are overestimated. 

\subsection*{Acknowledgements}
\noindent
I wish to thank M.\ Beneke, J.\ K\"uhn, T.\ Mannel, A.\ Martin, 
M.\ Seymour, and D.~Summers for interesting discussions.

\clearpage

\end{document}

%% file: uictabel1.tex
\begin{table}[htbp]
\centerline{
\begin{tabular}{  c c c c c c c | c c c c c c }
\hline
\multicolumn{2}{l}{$H=\psi(1^{--})$} 
 & \multicolumn{5}{ c  }{Colour-singlet} 
 & \multicolumn{6}{ c }{Colour-octet} 
\\ 
\hline
 $X=$
 & ${}^1S_0$  & ${}^3S_1$  & ${}^1P_1$  & ${}^3P_J$ & ${}^3D_{J'}$ 
 & ${}^1D_2$ 
 & ${}^1S_0$  & ${}^3S_1$  & ${}^1P_1$  & ${}^3P_J$ & ${}^3D_{J'}$  
 & ${}^1D_2$ 
\\ \hline
NRQCD & $v^8$ & $1$   & $v^8$ & $v^8$ & $v^8$ 
      & $v^{12}$
      & $v^4$ & $v^4$ & $v^8$ & $v^4$ & $v^8$ 
      & $v^{12}$
\\ 
CSM & $0$ & $1$ & $0$ & $0$ & $0$  
    & $0$
    & $0$ & $0$ & $0$ & $0$ & $0$  
    & $0$
\\ 
VSR1 & $v^2 \lambda^6$ & $1$   & $v^4 \lambda^4$ 
     & $v^2 \lambda^6$ & $v^4 \lambda^4$ 
     & $v^6 \lambda^6$
     & $v^2 \lambda^2$ & $\lambda^4$
     & $v^4 \lambda^4$ & $v^2 \lambda^2$ & $v^4 \lambda^4$
     & $v^6 \lambda^6$
\\ 
VSR2 & $v^7$ & $1$   & $v^7$ & $v^8$ & $v^8$ 
      & $v^{11}$
      & $v^3$ & $v^4$ & $v^7$ & $v^4$ & $v^8$ 
      & $v^{11}$
\\ 
CEM & $1$ & $1$ & $v^2$ & $v^2$ & $v^4$ 
    & $v^4$
    & $1$ & $1$ & $v^2$ & $v^2$ & $v^4$  
    & $v^4$
\\ 
\hline
\multicolumn{2}{l}{$H=\chi_{\Q J}(J^{++})$} 
  & \multicolumn{9}{ c }{} 
\\ \hline
NRQCD & $v^6$ & $v^6$ & $v^{10}$ & $v^2$ & $v^{10}$ 
      & $v^{10}$
      & $v^6$ & $v^2$ & $v^6$    & $v^6$ & $v^6$    
      & $v^{10}$
\\ 
CSM & $0$ & $0$ & $0$ & $v^2$ & $0$ 
    & $0$
    & $0$ & $0$ & $0$ & $0$   & $0$  
    & $0$
\\ 
VSR1 & $v^2 \lambda^4$ & $\lambda^6$ & $v^4 \lambda^6$
     & $v^2$ & $v^4  \lambda^6 $ 
     & $v^6 \lambda^4$ 
     & $v^2 \lambda^4$ & $\lambda^2$ & $v^4 \lambda^2$
     & $v^2 \lambda^4$ & $v^4 \lambda^2$ 
     & $v^6 \lambda^4$ 
\\ 
VSR2 & $v^5$ & $v^6$ & $v^{9}$ & $v^2$ & $v^{10}$ 
      & $v^{9}$
      & $v^5$ & $v^2$ & $v^5$    & $v^6$ & $v^6$    
      & $v^{9}$
\\ 
CEM & $1$ & $1$ & $v^2$ & $v^2$ & $v^4$ 
    & $v^4$
    & $1$ & $1$ & $v^2$ & $v^2$ & $v^4$ 
    & $v^4$
\\ \hline
\end{tabular}}
\caption[]{Relative scaling of $\langle {\cal O}_c^H(X) \rangle$.
\label{tab:scaling}
}
\end{table}%